\documentclass[aip,jcp,reprint]{revtex4-1}

\usepackage{float} 
\makeatletter
\let\newfloat\newfloat@ltx
\makeatother

\usepackage{algorithm}
\usepackage{algpseudocode}
\usepackage{graphicx}
\usepackage{amsmath,amssymb}
\usepackage{xcolor}
\usepackage[utf8]{inputenc} % allow utf-8 input
\usepackage[T1]{fontenc}    % use 8-bit T1 fonts
\usepackage{hyperref}       % hyperlinks
\usepackage{url}            % simple URL typesetting
\usepackage{booktabs}       % professional-quality tables
\usepackage{amsfonts}       % blackboard math symbols
\usepackage{amsmath}
\usepackage{nicefrac}       % compact symbols for 1/2, etc.
\usepackage{microtype}      % microtypography
\usepackage{graphicx}
\usepackage{lipsum}
\usepackage{comment}
\usepackage[caption=false]{subfig}
\usepackage{textcomp}
\usepackage{soul}           % highlight text \hl{...} and strike \st{...}
\usepackage{amsmath}
\usepackage{listings}
\usepackage{chngcntr}
\usepackage{enumitem}
\usepackage[table,xcdraw]{xcolor}
\usepackage{booktabs}
\usepackage{multirow}
\usepackage{hhline}
\usepackage{amssymb}
\usepackage{siunitx}
\usepackage[version=3]{mhchem}
\usepackage{euler-math}
\usepackage{bm}

\definecolor{codegreen}{rgb}{0,0.6,0}
\definecolor{codegray}{rgb}{0.5,0.5,0.5}
\definecolor{codepurple}{rgb}{0.58,0,0.82}
\definecolor{backcolour}{rgb}{0.95,0.95,0.92}

\lstdefinestyle{mystyle}{
    backgroundcolor=\color{backcolour},
    commentstyle=\color{codegreen},
    keywordstyle=\color{magenta},
    numberstyle=\tiny\color{codegray},
    stringstyle=\color{codepurple},
    basicstyle=\ttfamily\footnotesize,
    breakatwhitespace=false,
    breaklines=true,
    captionpos=b,
    keepspaces=true,
    numbers=left,
    numbersep=5pt,
    showspaces=false,
    showstringspaces=false,
    showtabs=false,
    tabsize=2
}
\let\oldtextbf=\textbf
\renewcommand*{\textbf}[1]{\ifmmode\mathbf{#1}\else\oldtextbf{#1}\fi}
\renewcommand*{\phi}[0]{\varphi}

\draft % marks overfull lines with a black rule on the right

\begin{document}

\title{Navigating committor landscape of biomolecules with a general pairwise interaction model.}

\author{Jintu Zhang}
\altaffiliation{These authors contributed equally: Jintu Zhang and Zichang Jin}
\affiliation{College of Pharmaceutical Sciences, Zhejiang University, Hangzhou 310058 Zhejiang, China}

\author{Zichang Jin}
\altaffiliation{These authors contributed equally: Jintu Zhang and Zichang Jin}
\affiliation{College of Pharmaceutical Sciences, Zhejiang University, Hangzhou 310058 Zhejiang, China}

\author{Huifeng Zhao}
\affiliation{College of Pharmaceutical Sciences, Zhejiang University, Hangzhou 310058 Zhejiang, China}

\author{Kai Zhu}
\affiliation{College of Pharmaceutical Sciences, Zhejiang University, Hangzhou 310058 Zhejiang, China}

\author{Bowei Zhao}
\affiliation{College of Pharmaceutical Sciences, Zhejiang University, Hangzhou 310058 Zhejiang, China}

\author{Xujun Zhang$^*$}
\email[]{xujunzhang@zju.edu.cn}
\affiliation{College of Pharmaceutical Sciences, Zhejiang University, Hangzhou 310058 Zhejiang, China}

\author{Peilin Kang$^*$}
\email[]{plkang@fudan.edu.cn}
\affiliation{State Key Laboratory of Porous Materials for Separation and Conversion, Collaborative Innovation Center of Chemistry for Energy Material, Shanghai Key Laboratory of Molecular Catalysis and Innovative Materials, MOE Key Laboratory of Computational Physical Science, Department of Chemistry, Fudan University, Shanghai, 200433, China}

\author{Tingjun Hou$^*$}
\email[]{tingjunhou@zju.edu.cn}
\affiliation{College of Pharmaceutical Sciences, Zhejiang University, Hangzhou 310058 Zhejiang, China}

\date{\today}

\begin{abstract}

Sampling rare conformation transitions between metastable states is a central challenge in atomistic simulations.
While the committor function serve as an ideal reaction coordinate for driving enhanced sampling, their high-dimensional inputs and complex functional forms limit the efficacy of standard feedforward neural networks in modeling them.
Inspired by recent breakthroughs in biomolecular structure prediction, we propose a novel committor learning framework grounded in the AlphaFold 3 paradigm.
By integrating a lightweight, differentiable atom-level embedding with a simplified Pairformer architecture, our method inherently captures intricate dynamical features of diverse biosystems without requiring specialized prior knowledge.
We demonstrate the superior expressiveness and accuracy of the proposed framework across multiple atomistic processes.
For the folding of the chignolin mini-protein, our model reveals the finer-grained structure of its transition state ensemble (TSE) and a detailed bifurcated reaction mechanism. 
Furthermore, for calixarene host–guest systems, we develop a unified committor model that elucidates how ligand substituents regulate the ratio between distinct binding pathways, offering new perspectives for structure-based drug design.
\end{abstract}

%\pacs{}% insert suggested PACS numbers in braces on next line

\maketitle %\maketitle must follow title, authors, abstract and \pacs

\section{Introduction}\label{sec:intro}

Atomistic simulations pose the problem of sampling macroscopic observables of atomic systems based on microscopic potential energy functions.
Rather than focusing solely on the thermodynamic properties of a single metastable state, the primary research interest has shifted to transitions between different free energy minima.
However, the presence of free energy barriers greatly hinders the occurrence of such transitions, which are thus termed rare events.
To bridge the gap between the waiting time of rare events and affordable simulation timescale, since the seminal work by \citeauthor{torrie1977}, enormous efforts have been put into the development of enhanced sampling methods.~\cite{henin2022enhanced,kai2026review}

Recently, \citeauthor{kang2024committor} have introduced a committor-based enhanced sampling method to accelerate the sampling not only of the metastable states but also of the transition states (TSs).~\cite{kang2024committor,trizio2025committor,trizio2026pipe}
Given two configuration space regions A and B, the committor $q(\symbf{x})$ is a function that indicates the probability of a trajectory originating from conformation $\symbf{x}$ first entering B without reaching A.~\cite{weinan2010tpt}
It has been shown that the committor function is an ideal reaction coordinate (RC),~\cite{ma2005automatic} and thus offers a comprehensive understanding of transition processes.
Providing both accurate free energy estimation and an explanation of the reaction mechanism, such a method has found use in a variety of real-world systems.~\cite{kang2024committor,trizio2025committor,deng2026role,das2025machine,alessandro2026lcc,vidal2026enzyme}

Due to the high dimensionality and nonlinearity of the committor function, \citeauthor{kang2024committor} proposed leveraging neural networks (NNs) to learn the committor, as do many other established methods.~\cite{khoo2019solving,rotskoff2022active,mitchell2024committor,chen2023discovering,jung2023machine,megias2025iterative,breebaart2026understanding}
In practice, the network receives as input a set of descriptors $\symbf{d}$ that are invariant under symmetry operations of atomic systems: $q(\symbf{x})=q_{\theta}(\symbf{d}(\symbf{x}))$.
Following the variational principle derived from the backward Kolmogorov equation,~\cite{weinan2010tpt,kolmogorov1931analytischen} the network weights ($\theta$) are optimized to minimize the functional:
\begin{equation}
\mathcal{K}[q(\symbf{x})]=\left\langle\left|\nabla_{\symbf{u}}q(\symbf{x})\right|^2\right\rangle_{U(\symbf{x})},
\label{equ:functional_1}
\end{equation}
where the ensemble average $\langle\cdot\rangle_{U(\symbf{x})}$ is taken over the Boltzmann distribution governed by the potential $U(\symbf{x})$, and $\symbf{u}$ denotes the mass-weighted coordinates.

\begin{figure*}[th!]
\centering
\includegraphics[width=\textwidth]{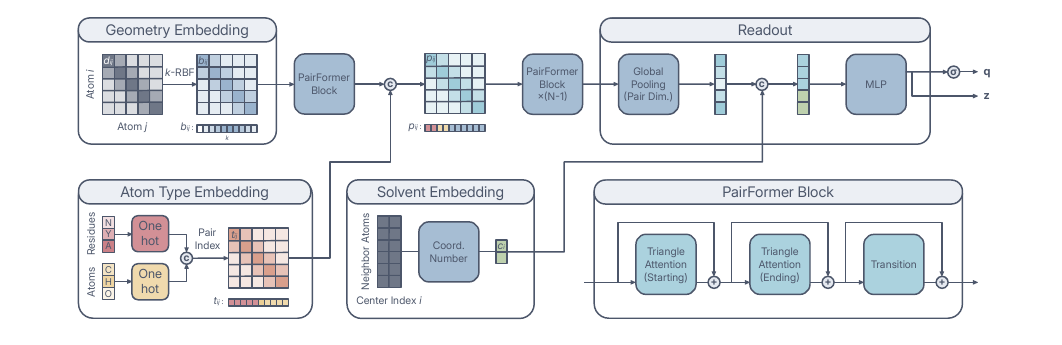}
\caption{Overall architecture of the Pairformer-based committor model. The model processes inputs through three embedding modules: geometry (top-left panel), atom type (bottom-left panel), and solvent (bottom-middle panel). A primary Pairformer block first extracts geometric features, which are then concatenated with atom type features and processed by the main Pairformer stack. Subsequently, pairwise features are pooled into a conformation-level tensor, concatenated with the solvent embedding, and projected to a one-dimensional scalar output. The internal structure of the simplified Pairformer block is illustrated in the bottom-right panel.}
\label{fig:0}
\end{figure*}

Nevertheless, in general, effectiveness of variational methods depends on their choice of basis functions.
In the context of committor learning, such a limitation requires descriptors closely associated with the transition process to be selected as inputs of the network.~\cite{kang2024committor,trizio2025committor}
To mitigate the burden on users, advanced machine learning techniques, e.g., geometric Graph Neural Network (GNN), have been adopted to parametrize the committor directly from atomic coordinates $\symbf{x}$.~\cite{zhang2024gnncv,kang2026committor,contreras2026committor,esders2025analyzing}
For the sake of efficiency, GNNs generally leverage the locality hypothesis of atomic systems, which assumes that interactions between atoms separated by long distances are negligible.~\cite{duval2023hitchhiker}
Conversely, even if the potentials are localized, committor functions of large-scale spatial transitions have to directly respond to long-range interatomic motions.~\cite{kang2024committor,trizio2025committor,jung2023machine}
To this end, \citeauthor{kang2026committor} presented a multiscale GNN method, the dual-cutoff GNN, which captures long-range behaviors by using a much larger interaction range for the subset of atoms dominating the large-scale motions.~\cite{kang2026committor}

The recent success of the AlphaFold series in protein structure prediction has shed new light on representation learning methods for biomolecules.~\cite{af1,af2,af3}
In its most advanced form, AlphaFold 3 (AF3) employs a hierarchical representation scheme that aggregates detailed atom-level information into coarse-grained token representations, which are subsequently processed by the Pairformer architecture~\cite{af3} for accurate structural reconstruction.
As the core component of AF3, Pairformer utilizes the triangle attention mechanism to automatically model intricate relationships between structural units (e.g., protein residues or ligand atoms).
Moreover, as a dense method, Pairformer aggregates information from every possible token pair, thus inherently accounting for long-range correlations across distinct structural domains.
Consequently, Pairformer is a promising candidate to obtain expressive committor functions for biomolecules.

Here, we introduce a Pairformer-based committor learning framework that extends the expressive power of Pairformer architectures to the identification of reaction coordinates and transition mechanisms in complex biological systems.
The resulting framework is illustrated in Fig.~\ref{fig:0}.
To validate the accuracy and applicability of the Pairformer-based approach, we conducted simulations of several atomistic processes.
Firstly, in alignment with the tradition of the enhanced sampling community, we tested the approach on the conformation transition of alanine dipeptide in a vacuum.
Secondly, we studied the folding of the chignolin mini protein in bulk water.
Enabled by the high resolution of the proposed method, finer-grained structures of the transition state ensemble (TSE) are revealed, along with a more detailed bifurcation reaction mechanism.
Finally, we investigated the binding processes of a series of aromatic ligands with a calixarene host in bulk water.
Despite their simplicity, such a set of model systems represents the core mechanics of drug-protein interactions.
Leveraging the proposed method, a unified committor model was trained to simultaneously account for all binding processes, permitting a consistent treatment of each ligand.
Through the resulting committor model, we elucidated the regulatory effect of ligand substituents on the ratio between different binding pathways, providing a new perspective on structure-based drug design.

\section{Results}

\subsection{Alanine dipeptide}

\begin{figure}[th!]
\centering
\includegraphics[width=\linewidth]{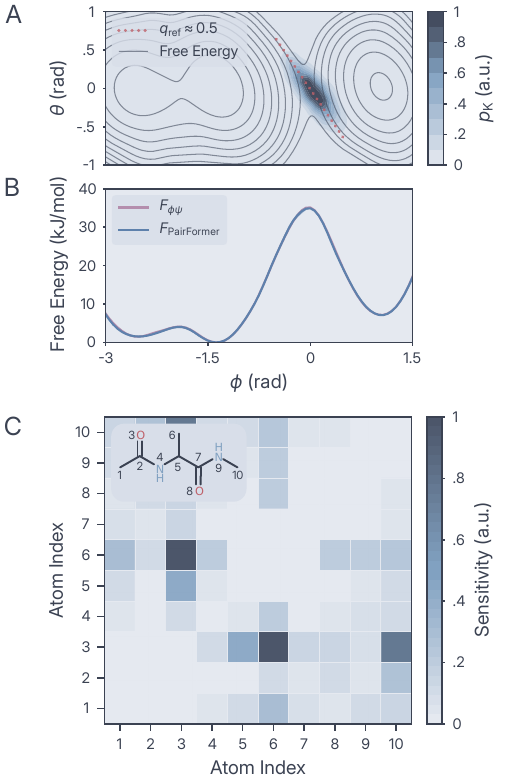}
\caption{Alanine dipeptide. (A) Contour plot of the unnormalized Kolmogorov distribution $e^{-\beta U_{\mathcal{K}}(\mathbf{x})}$. Gray isolines indicate the free energy landscape, and the red dotted line shows the relationship between $\phi$ and $\theta$ for TSE configurations as proposed by \citeauthor{bolhuis2000committor}. (B) Free energy profile projected along the $\phi$ coordinate. The blue and purple curves represent results from the $z$-based and $\phi\psi$-based OPES simulations, respectively. (C) Sensitivity scores for each atom pair.}
\label{fig:1}
\end{figure}

Following the established convention in the enhanced sampling community, we demonstrate the effectiveness of our method on the conformational equilibrium between the C7\textsubscript{eq} and C7\textsubscript{ax} metastable states of alanine dipeptide in a vacuum.
Among the vast literature studying this system, backbone torsional angles $\phi$ and $\psi$ are generally used to drive free energy calculations, whereas $\phi$ and $\theta$ are applied for parameterizing the accurate committor.~\cite{bolhuis2000committor,kang2024committor}

To train the model, we adopted the dataset reported in ref.~\citenum{kang2026committor}, and used the 10 non-hydrogen atoms of the molecule to build the inputs.
We found that the variational loss converged to the same order of magnitude within the same number of epochs compared to the existing result (see Fig.~S1 of the Supplementary Information), demonstrating the accuracy of the model.
We then carried out a simulation using the Kolmogorov bias to sample the TSE (Fig.~\ref{fig:1}A), yielding results that are also in agreement with the prior study.~\cite{trizio2025committor}
Besides, OPES simulations based on the $z$ component of the model outputs generate a free energy profile identical to the one generated by the OPES simulation driven by $\phi\psi$ (Fig.~\ref{fig:1}B).
In addition, sensitivity analysis performed on the TSE conformations suggests that distances between atom pair 3-5 and 3-6 contribute most to the committor (Fig.~\ref{fig:1}C).
The same conclusion was also drawn in the previous work.~\cite{kang2024committor}

\subsection{Chignolin mini-protein folding}

\begin{figure*}[th!]
\centering
\includegraphics[width=\textwidth]{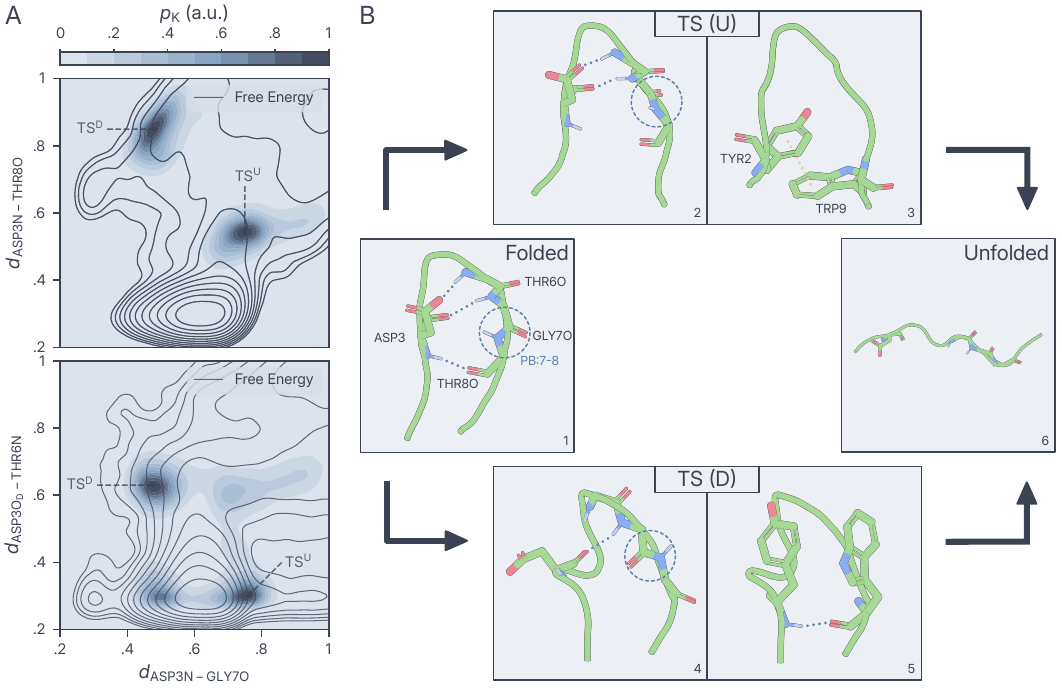}
\caption{Chignolin folding. Contour plots of the unnormalized Kolmogorov distribution $e^{-\beta U_{\mathcal{K}}(\mathbf{x})}$ projected onto subspaces spanned by different interatomic distances. Gray isolines indicate the free energy landscape. (B) Representative structural snapshots of the folded, unfolded, and transition states. In panels 1, 2, 4, and 6, the heavy atoms of ASP3 and the backbone atoms of THR6, GLY7, and THR8 are shown in a licorice representation, with the peptide bond between GLY7 and THR8 highlighted by a dashed circle. In panels 3 and 5, the heavy atoms of TYR2 and TRP9 are displayed as licorice. Across all panels, hydrogen bonds and $\pi-\pi$ stacking interactions are denoted by blue and yellow dashed lines, respectively.}
\label{fig:2}
\end{figure*}

We applied our method to investigate the folding process of the chignolin mini-protein in bulk water.
As a 10-residue peptide, chignolin spontaneously folds into a stable $\beta$-hairpin structure in an aqueous environment.
The progress of this folding event, as previously demonstrated, is characterized by the formation of a hydrogen-bond network among residues ASP3, THR6, GLY7, and THR8.~\cite{trizio2025committor,megias2025iterative}

To train the model, we adopted the dataset reported in ref.~\citenum{trizio2025committor}, and used the 40 backbone atoms of the peptide to build the inputs.
Using this model, we first performed Kolmogorov bias-based simulations to sample the TSE (Fig.~\ref{fig:2}A).
In agreement with prior studies, two distinct sets of TS conformations are found, namely, TS\textsuperscript{U} and TS\textsuperscript{D}.~\cite{kang2024committor,trizio2025committor}
As illustrated in Fig.~\ref{fig:2}A, starting from the folded state, reaction trajectories passing through TS\textsuperscript{U} are initiated by the breaking of the ASP3N-THR8O hydrogen bond and a modest increase in the ASP3N-GLY7O distance, while the hydrogen bond ASP3O\textsubscript{D}-THR6N tends to remain intact.
This mechanism indicates a direct unzipping of the peptide termini, with the backbone torsions essentially unaffected.
The representative structure drawn from TS\textsuperscript{U} also supports this hypothesis, as shown in Fig.~\ref{fig:2}B.
On the other hand, as the system approaches TS\textsuperscript{D}, the ASP3N-THR8O distance increases substantially, while the ASP3N-GLY7O distance decreases slightly.
Since GLY7O and THR8O are spatially proximal, the opposing trends in their respective distances to ASP3N imply a backbone rearrangement (Fig.~\ref{fig:2}A).
Fig.~\ref{fig:2}B shows the representative structure of TS\textsuperscript{D}, and as anticipated, compared to TS\textsuperscript{U}, the GLY7-THR8 peptide bond rotates by approximately 180$^\circ$.
Furthermore, the rather low formation probability of the ASP3O\textsubscript{D}-THR6N hydrogen bond in TS\textsuperscript{D} suggests a twisting of the N-terminus of the peptide.
Thus, TS\textsuperscript{D} corresponds to the reaction pathway in which the breaking of the hydrogen-bond network is initiated by the deformation of the backbone structure.

The different backbone conformations of TS\textsuperscript{U} and TS\textsuperscript{D} further affect the orientations of TYR2 and TRP9 (Fig.~\ref{fig:2}B).
Due to the folded-like backbone structure, the sidechains of TYR2 and TRP9 are likely to form $\pi-\pi$ stacking in TS\textsuperscript{U}.
Conversely, $\pi-\pi$ stacking is less likely to form in TS\textsuperscript{D} because of the tension of the twisted backbone.
To compensate, the flipped orientations of the two residues allow an additional TYR2N–TRP9O hydrogen bond to form, rendering this transition state also energetically favorable.

We then performed the sensitivity analysis using the TS conformations ($-0.5 < z < 0.5$), demonstrating that interatomic distances between residues ASP3, THR6, GLY7, and THR8 contribute significantly to the committor (Fig.~S2A).
Utilizing the 22 most important distances identified by the sensitivity analysis, we performed a LASSO regression, yielding a linear approximation of the committor based on seven distances (see Section I.B.3 of the Supplementary Information for further details).
Following the protocol described in ref.~\citenum{ray2025lasso}, we performed OPES simulations by biasing the LASSO CV.
We found that the free energy difference between the folded and unfolded states converged within 2 \si{\us} (Fig.~S2B), which is slightly faster than originally reported.~\cite{ray2025lasso}

\subsection{Calixarene host-guest systems}

\begin{figure*}[th!]
\centering
\includegraphics[width=\textwidth]{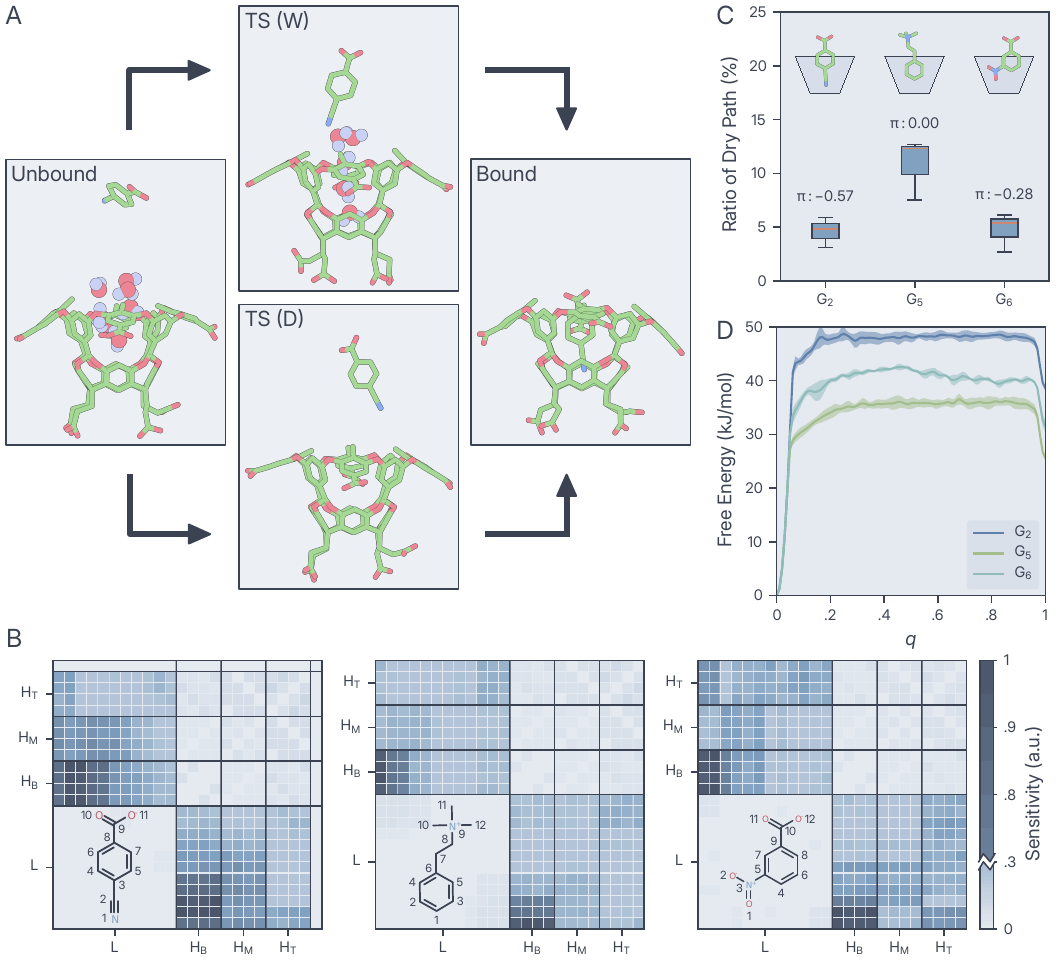}
\caption{Calixarene host-guest systems. (A) Representative snapshots of the unbound, bound, and transition states of the calixarene–G\textsubscript{2} complex. (B) Sensitivity scores for each atom pair of all studied systems. (C) Dry-to-wet pathway ratios for each system; average values and standard errors were estimated from three independent simulations. The Hansch hydrophobic constants ($\pi$) of various lower-region substituents~\cite{hansch1995exploring} are displayed above the respective plots along with schematic representations of the bound-state conformations. (D) Free energy profiles for each system projected along $q$. Solid lines denote the average values obtained from three independent simulations, while the shaded regions represent the standard deviations.}
\label{fig:3}
\end{figure*}

As the final test for our method, we investigated the binding processes of a series of aromatic ligands with a calixarene host in bulk water.
Being proposed in the \verb|SAMPL5| challenge,~\cite{yin2017overview} the G\textsubscript{2} (4-cyanobenzoic acid), G\textsubscript{5} (N,N,N-trimethyl-2-phenylethan-1-aminium) and G\textsubscript{6} (3-nitrobenzoic acid) ligands were included in our test.
Although structurally simple, such systems maintain the essential features of drug-protein interactions.
Previous studies have shown that the binding process proceeds via two distinct pathways: a ``wet'' path, which first leads to a semi-bound state B\textsubscript{w} with water molecules retained in the pocket before reaching the fully-bound state B\textsubscript{d}, and a ``dry'' path, where the ligand enters the binding pocket directly.~\cite{trizio2025committor,trizio2026pipe,kang2026committor}
The respective structures of the TSs involved in the different pathways are shown in Fig.~\ref{fig:3}A.
Despite the similar ligand positions, the binding pocket is solvated in the wet TS (TS\textsuperscript{W}), but becomes completely desolvated in the dry TS (TS\textsuperscript{D}).
Additionally, the wet path has been shown to be dominant even if it involves more intricate dynamics.

Enabled by the ability of Pairformer to describe diverse systems concurrently, our study goes beyond previous works by accounting for the binding processes of different ligands in one unified committor model.
To build inputs for the model, we used all non-hydrogen ligand atoms and three groups of host atoms, which, according to their relative locations within the host structure, were classified as bottom (H\textsubscript{B}), middle (H\textsubscript{M}) and top (H\textsubscript{T}) atoms, as shown in Fig.~S3A.
Besides, based on the geometric centers of these three atom groups, a solvent embedding was incorporated into the model to capture the local solvent structure information.
Having the model defined, we started the self-consistent fitting iterations from unbiased simulations in the dry bound state (B\textsubscript{d}), the unbound state, and OPES simulations driven by two suboptimal, however intuitive, CVs.
Due to the expressivity of our approach, we found the committor to be accurate after two iterations (see Section I.C.2 of the Supplementary Information for details).

We first examined the knowledge acquired by the model using sensitivity analysis.
As expected, host-ligand distances contribute most significantly to the committor, while intra-ligand and intra-host distances play almost no role (Fig.~\ref{fig:3}B).
Interestingly, the model tends to focus more on the lower-region substituents of the ligands, specifically the functional groups deeply buried within the binding pocket (the cyano group of G\textsubscript{2}, the non-substituted \textit{p}-carbon of G\textsubscript{5} and the nitro group of G\textsubscript{6}).
Since the carboxyl groups of G\textsubscript{2} and G\textsubscript{6} and the quaternary ammonium group of G\textsubscript{5} offer comparable hydrophilicity,~\cite{hansch1995exploring} the lower-region substituents serve to fine-tune the overall hydrophilicity and binding behavior of the ligands.
To gain insight into how the model differentiates the binding behaviors of different ligands, we carried out Kolmogorov bias-based simulations to sample the TSE of each system (Fig.~S3B).
Remarkably, we observed that the dry-to-wet pathway ratio varies across different systems: the G\textsubscript{2} and G\textsubscript{6} systems exhibit a ratio of approximately 5\%, whereas the G\textsubscript{5} system exhibits a ratio of about 11\%, demonstrating that G\textsubscript{5} has a higher propensity for binding through the dry path (Fig.~\ref{fig:3}C).
According to the Quantitative Structure-Activity Relationship (QSAR) theory, the cyano group in G\textsubscript{2} and the nitro group in G\textsubscript{6} are both hydrophilic (with a negative Hansch hydrophobic constant, $\pi < 0$), whereas the hydrogen atom linked with the \textit{p}-carbon of G\textsubscript{5} is neutral ($\pi = 0$).
Thus, the simulation results align well with chemical intuition, clearly demonstrating the modulatory effect of distinct functional groups on the binding mechanism.

Finally, leveraging the committor model, we performed OPES simulations and calculated the free energy profile of each system (Fig.~\ref{fig:3}D).
We found that these free energies are in good agreement with the previously reported reference values (Fig.~S3C),~\cite{trizio2025committor} and their ranking successfully reproduces the experimental trends.~\cite{yin2017overview}

\section{Discussions}

In this work, we have presented a committor fitting framework based on the Pairformer architecture.
By leveraging the exactitude of the variational principle and the expressive power of Pairformer, the proposed framework precisely captures the essential dynamical features of various atomistic processes, representing a synergistic combination of enhanced sampling methods and cutting-edge biomolecular structure prediction techniques.
We applied our method to investigate three biological systems: alanine dipeptide, chignolin mini-protein, and calixarene host–guest complexes.
In all cases, our framework produced accurate committor models and yielded deep insights into the underlying dynamics.
Specifically, for the calixarene systems, we demonstrated the regulatory effect of ligand substituents on the ratio between wet and dry binding pathways, which aligns well with traditional QSAR-based drug design methodologies.

Nonetheless, further improvements are still warranted.
Although Pairformer is remarkably expressive, the associated computational cost remains considerable for large systems.
Thus, it may be beneficial to incorporate lower-scaling methods, such as the pair-biased Transformer employed in La-Proteina~\cite{geffner2026laproteina}.
Alternatively, the model could also benefit from a coarser-grained tokenization scheme for biomolecular structures.
Besides, grounded in the AF3 paradigm, we believe that the direct generation of accurate TSE conformations for diverse biosystems should also be feasible, thereby leading to the next era of enhanced sampling.

\section{Methods}

\subsection{Background}

Following the variational principle derived from the backward Kolmogorov equation, a function $q(\symbf{x})$ that minimizes the functional:
\begin{equation}
\mathcal{K}[q(\symbf{x})]=\left\langle\left|\nabla_{\symbf{u}}q(\symbf{x})\right|^2\right\rangle_{U(\symbf{x})}
\label{equ:functional_2}
\end{equation}
subject to the boundary conditions $q\left(\symbf{x}_A\right)=0$ and $q\left(\symbf{x}_B\right)=1$ is the committor function of the $A\to B$ process of an atomistic system that undergoes the overdamped Langevin dynamics governed by the potential $U(\symbf{x})$.~\cite{kolmogorov1931analytischen,weinan2010tpt}
However, fundamentally, the discrete evaluation of the ensemble average in Eq.~\ref{equ:functional_2} requires weights associated with the sample conformations under an equilibrium Boltzmann distribution, which in turn relies on equilibrium sampling of the configuration space.
To build committor models in an ``ab initio'' fashion, \citeauthor{kang2024committor} presented a self-consistent iterative workflow.~\cite{kang2024committor,trizio2025committor}
The workflow begins with a trial committor model trained solely on boundary conformations, which are obtained from unbiased sampling of the two boundary states.
Bootstrapped by the trial model, each step of the workflow then comprises a sampling stage and a training stage: the previously fitted committor model is utilized to drive enhanced sampling simulations, thereby obtaining more conformations and more accurate weights; subsequently, these newly generated data are used to retrain and refine the model.

During the sampling stage of the workflow, two types of biasing potentials can be applied to flatten energy barriers, thus promoting the equilibrium sampling.
The first one is the so-called Kolmogorov bias that takes the form:
\begin{equation}
V_{\mathcal{K}}(\symbf{x})=-\frac{\lambda}{\beta}\symsf{log}\left|\nabla_{\symbf{u}}q(\symbf{x})\right|^2,
\label{equ:k_bias}
\end{equation}
where $\beta$ is the inverse temperature and $\lambda$ a factor that controls the strength of the potential.
As demonstrated in ref.\citenum{kang2024committor}, such a potential essentially turns the transition state into another metastable state, enabling extensive sampling of the transition region.
The second type of biasing potential is a metadynamics-like one provided by the on-the-fly probability enhanced sampling (OPES) method~\cite{invernizzi2020opes}, which further improves the ergodicity of the sampling.
To address the vanishing gradient problem exhibited by the committor function inside metastable basins when acting as a CV, the committor is, in fact, expressed as a nonlinear transformation of an auxiliary variable $z$: $q(x) = \sigma(z(\symbf{x}))$, where $\sigma(z) = 1/(1+e^{-pz})$ is the sigmoid function.
The $z$ variable, which encodes the same information as the committor but is much smoother over the configuration space, is then used as the CV for applying the OPES bias.~\cite{trizio2025committor}

Beside accelerating the sampling, one crucial application of the Kolmogorov bias is to properly define the TSE.
In such an ensemble, the weights of the conformations obey the Kolmogorov distribution:
\begin{equation}
p_{\mathcal{K}}(\symbf{x})=\frac{e^{-\beta U_{\mathcal{K}}(\symbf{x})}}{{\mathcal{Z}_{\mathcal{K}}}},\ U_{\mathcal{K}}(\symbf{x}) = V_{\mathcal{K}}(\symbf{x}) + U(\symbf{x}),
\label{equ:p_k}
\end{equation}
where ${\mathcal{Z}_{\mathcal{K}}}$ is a normalization constant.~\cite{kang2024committor}
According to the variational principle, the above definition measures contributions to the transition rate ($\nu_R$) made by trajectories passing through conformation $\symbf{x}$, when the strength factor $\lambda$ is about 1:~\cite{weinan2010tpt}
\begin{equation}
\nu_R \propto \int d \symbf{x} p_{\mathcal{K}}(\symbf{x})
\label{equ:rate}
\end{equation}
By sampling $p_{\mathcal{K}}(\symbf{x})$, one obtains not only structural, but also thermodynamic information about the TSE.

\subsection{The machine learning committor function}

As outlined in the introduction, we model the committor function using Pairformer, a representation learning architecture originally introduced in AF3.
As a biomolecular structure prediction model, AF3 constructs inputs for Pairformer by leveraging sequence information, initial residue geometries, and spatial features extracted from homologous templates.
To capture the highly intricate sequence-structure relationship, AF3 also employs a deep stack of Pairformer blocks to process these input representations~\cite{af3}.
In contrast, the committor model fundamentally acts as a mapping from the configuration space (and potentially the atom type space) to scalar values.
As a result, the complicated embedding frameworks adopted by AF3 need to be modified to lower their computational costs and obey the general requirements of a CV model.
Thus, in this work, we introduced a lightweight embedding framework to construct inputs for Pairformer, and a GNN-inspired readout module to yield the final scalar committor values.
Meanwhile, we largely simplified the original Pairformer architecture to further accelerate training and inference.
With these modifications incorporated, the model was then optimized using the variational committor loss function, as described in ref.~\citenum{kang2024committor,trizio2025committor,kang2026committor}

While these model components are closely interrelated, for clarity, the remainder of this section first details the loss function, followed by the simplification of Pairformer and the design of the embedding and readout modules.

\subsubsection{Loss function}

The loss function consists of three terms:
\begin{equation}
\mathcal{L} = \symsf{log} \mathcal{L}_{v} + \alpha_1 \mathcal{L}_{b} + \alpha_2 \mathcal{L}_{r}
\label{equ:loss_t}
\end{equation}
where the hyperparameters $\alpha_1$ and $\alpha_2$ are the weights of them.
In the first term, $\mathcal{L}_v$ is the functional $\mathcal{K}[q(\symbf{x})]$ estimated on a dataset containing $N_v$ conformations:
\begin{equation}
\mathcal{L}_v = \frac{1}{N_v} \sum_{i=1}^{N_v} w_i |\nabla_{\symbf{u}} q(\symbf{x}_i)|^2,
\label{equ:loss_v}
\end{equation}
where $w_i$ is the statistical weight of the conformation $\symbf{x}_i$.~\cite{trizio2025committor}
Following the methodology presented in ref.~\citenum{deng2026role,kang2026committor}, rather than directly using $\mathcal{L}_v$ as the variational loss term, the logarithm of $\mathcal{L}_v$ is adopted to improve convergence behavior.
Due to the variational bound, $\mathcal{L}_v$ is strictly greater than 0 yet can be arbitrarily small.
Consequently, taking its logarithm facilitates more thorough training and mitigates numerical instability.

The second term, $\mathcal{L}_b$, imposes the boundary conditions $q(\symbf{x}_A) = 0$ and $q(\symbf{x}_B) = 1$ on the model:
\begin{equation}
\mathcal{L}_b = \frac{1}{N_A} \sum_{\symbf{x} \in A} (q(\symbf{x}) - 0)^2 + \frac{1}{N_B} \sum_{\symbf{x} \in B} (q(\symbf{x}) - 1)^2.
\label{eq:loss_b}
\end{equation}
In practice, $\mathcal{L}_b$ is evaluated on a labeled dataset that contains $N_A$ and $N_B$ conformations belonging to states A and B respectively.
Finally, a regularization term $\mathcal{L}_{r}$ is introduced to restrain the accessible range of $z$ and to make optimizations more balanced, as implemented in ref.~\citenum{kang2026committor}:
\begin{equation}
\mathcal{L}_{r}=\sum_{i=1}^{N}\symbb{1}(\delta z_i)(\delta z_i)^2/\sum_{i=1}^{N}\symbb{1}(\delta z_i),\ \delta z_i=|z_i|-z_r,
\label{eq:loss_r}
\end{equation}
where $z_r$ is a user set threshold value and $\symbb{1}(\delta z_i)$ an indicator function that takes the value of 1 if $\delta z_i > 0$, and 0 otherwise.
To ensure its effectiveness, $\mathcal{L}_{r}$ is evaluated on all conformations present in the training set.

\subsubsection{Pairformer architecture}

The original Pairformer model, which is composed of 48 Pairformer blocks, takes as inputs both the token-level single representations $\symbf{s} \in \symbb{R}^{n\times c}$ and pair representations $\symbf{p} \in \symbb{R}^{n\times n\times c}$ (where $n$ is the number of tokens and $c$ the size of each token).
Within each Pairformer block, pair representations undergo a sequential transformation through two triangle multiplicative updates, two triangle self-attentions, and a transition module.~\cite{af3}
In the four triangle modules, following the triangle inequality on distances, the representation of the token pair $ij$ ($\symbf{p}_{ij}$) is updated by aggregating information from neighboring pairs that are directly attached to the pair $ij$ (through token $i$ or $j$).~\cite{af2}
While multiplicative update combines information from all neighboring pairs equally, self-attention applies a multihead attention mechanism~\cite{vaswani2017attention} to achieve more dynamic information updating.
Afterward, an element-wise transition module processes the pair representations, injecting greater non-linearity into the network.
Finally, with the fully updated pair representations, the single representations are updated by a pair-bias module.~\cite{af3}

To improve computational efficiency, we made several modifications to the original Pairformer model.
First, since the committor learning task does not utilize sequence data directly, we removed the whole single representation channel, and absorbed the atom type information in pair representations.
As single representations do not contribute to the update of pair representations, their removal preserves overall model accuracy.
Besides, as pointed out in ref.~\citenum{af2,geffner2025laproteina}, although the triangle multiplicative updates improve model accuracy by a few percentage points, they introduce substantial computational cost.
Given these considerations, we also removed these update modules.
Owing to the same triangle inequality embodied by the self-attention module, the simplified model retains the capability of capturing complex interatomic geometric relationships.
We show the resulting model architecture in the bottom-right panel of Fig.~\ref{fig:0}, and provide detailed equations in the Supplementary Information (Algorithm 1 to 4).

\subsubsection{Embedding}

To capture as much information of the system without loss of generality, we employed three embedding modules to construct inputs for the model, namely, the geometry embedding, the atom type embedding and the solvent embedding.
Unlike AF3, we directly used atom-level pair representations as model inputs, i.e., treating each atom as an individual token, to learn the fine-grained dynamical behavior of various systems.
Nevertheless, the tokenization scheme can still be easily switched to a coarse-grained one if the system contains numerous atoms, thanks to the universality of Pairformer.

AF3 models the geometric information of homologous template structures using a distogram method, which discretizes pairwise distances into ranges via one-hot encoding.~\cite{af3}
Due to its discrete nature, the distogram is non-differentiable, which precludes its use in committor models, since the variational loss function (Eq.~\ref{equ:loss_v}) requires gradients of the committor with respect to atomic coordinates.
To circumvent this limitation, we replaced the distogram with its continuous counterpart, the Gaussian radial basis function (RBF), which is widely applied by geometric GNNs.~\cite{duval2023hitchhiker,gasteiger2022}
Therefore, the geometry embedding of our model is the pairwise distances expanded by RBFs (top-left panel of Fig.~\ref{fig:0}), dimensionally matching with the ``distogram-linear'' approach in AF3.
Subsequently, the outputs of the RBFs ($\symbf{b} \in \symbb{R}^{n\times n\times c}$) first go through a single Pairformer block to encode the pure geometric information.
This design ensures that even when a limited number of Pairformer blocks are used, the model remains focused on the geometry of the input conformation.

Meanwhile, we adopted a standard one-hot encoding scheme for the atom-type embedding.
As shown in the bottom-left panel of Fig.~\ref{fig:0}, both element and residue types are combined to construct single-atom tokens.
These tokens are then concatenated with the outputs of the primary geometry-focused Pairformer block to synthesize the inputs ($\symbf{p}$) for the main Pairformer stack.

While the above embedding framework is sufficient for modeling the committor function of most unimolecular processes, an important class of biological phenomena, host-ligand binding, is generally regulated by environmental water molecules.~\cite{rizzi2021water,ansari2022water}
However, due to their high mobility and potential large number, AF3 does not explicitly account for waters.
Consequently, it is worthwhile to integrate an additional solvent embedding into the model, thereby making it responsive to solvent structures.
To this end, we adopted the concept of water coordination number (CN) from ref.~\citenum{rizzi2021water,ansari2022water,trizio2025committor}, which quantifies the degree of solvation at given spatial points.
Because the CN is a smooth function of atomic positions, it can serve as the solvent embedding of committor models, as demonstrated in ref.~\citenum{trizio2025committor}.
For convenience, we defined the spatial points as the geometric centers of different sets of host atoms, rather than virtual atoms with fixed positions as proposed in ref.~\citenum{rizzi2021water}.
As shown in the bottom-middle panel of Fig.~\ref{fig:0}, the solvent embedding is a conformation-level quantity that does not necessarily require processing by the Pairformer stack.
Therefore, we bypassed the Pairformer stack and feed this embedding directly into the readout module as a parallel feature alongside the Pairformer output.
We provide the detailed definition of the CN in the Supplementary Information (Algorithm 5).

\subsubsection{Readout}

Akin to GNNs, we employed a mean pooling operation to aggregate token-pair features from the Pairformer stack output, as shown in the top-right panel of Fig.~\ref{fig:0}.
This conformation-level feature tensor of dimension $c$ is then concatenated with the solvent embedding and processed by a feedforward network.
The 1-dimensional output of this readout network is the $z$ variable, which is then transformed into the committor $q$ via a sigmoid function.
We provide the complete Pairformer-committor architecture in the Supplementary Information (Algorithm 6).

\subsection{Interpreting the model}

To distinguish the most relevant atom pairs in the learned committor model, a gradient-based sensitivity analysis can be applied.
As presented in ref.~\citenum{kang2024committor}, the symmetrized sensitivity $\symbb{s}$, or importance of a given pair, is defined as its averaged contribution to the model output over a dataset of $N_c$ conformations:
\begin{equation}
\symbb{s}_{ij}=\symbb{s}_{ji}=\frac{\left(\tilde{\symbb{s}}_{ij}+\tilde{\symbb{s}}_{ji}\right)}{2},\ \tilde{\symbb{s}}_{ij} = \frac{1}{N_c}\sum_{k=1}^{N_c}\left|\frac{\partial q_{\theta}\left({\symbf{d}^k}\right)}{\partial{\symbf{d}^k_{ij}}}\right|\Delta_k\left(\symbf{d}_{ij}\right),
\label{eq:sensitivity}
\end{equation}
where $\symbf{d}^k$ and $\Delta_k\left(\symbf{d}_{ij}\right)$ represent the distance matrix of the conformation $k$ and the distance range of the pair $ij$ over the dataset, respectively.

While the above pair-level sensitivity analysis provides insights into the importance of individual pairs, a linear approximation of the model is often desirable.
Following ref.~\citenum{novelli2022lasso,zhang2024gnncv}, we used the Least Absolute Shrinkage and Selection Operator (LASSO) regression~\cite{tibshirani1996regression} to obtain such symbolic expressions:
\begin{equation}
f_{\symbf{w}}(\symbf{d}) = \sum_{ij} \symbf{w}_{ij}\ \symbf{d}_{ij},
\label{eqn:lasso}
\end{equation}
where $\symbf{w}_{ij}$ is the combination coefficient.
As a sparse method, LASSO regression fulfills the sparsity condition, which means that only a few coefficients are non-zero.
In practice, this is achieved by optimizing the LASSO loss function, which imposes a penalty to the magnitude of the coefficients via $L_1$ norm:
\begin{equation}
\mathcal{L}_{\mathrm{LASSO}} = \frac{1}{N_c} \sum_{k=1} ^{N_c} \left|f_{\symbf{w}}\left(\symbf{d}^k\right)-z_{\theta}\left(\symbf{d}^k\right)\right|^2+ \alpha\left|\symbf{w}\right|_1,
\label{eqn:lasso_loss}
\end{equation}
where the regularization strength $\alpha$ measures the level of sparsity, i.e., how many coefficients are different from zero.

However, due to the limited expressive power of linear models, applying LASSO regression directly to large distance matrices might yield degenerate expressions.
Therefore, we employed a hybrid strategy to obtain linear approximations of the committor models: we first used sensitivity analysis to select approximately the top 10\% most important pairwise distances and subsequently applied LASSO regression to this subset of features with high importance.
By focusing exclusively on such features, this workflow guarantees a physically meaningful linear model while retaining essential predictive power.

\section*{Code and Data Availability} \label{sec:code_avail}
The code for training the Pairformer-based committor model is available through a fork of the \texttt{mlcolvar} library~\cite{bonati2023mlcolvar}:  \hyperlink{https://github.com/jintuzhang/mlcolvar}{https://github.com/jintuzhang/mlcolvar}. Training and simulation data and inputs are available on \hyperlink{https://github.com/jintuzhang/pairformer\_committor}{https://github.com/jintuzhang/pairformer\_committor}.

\section*{Supplementary Information} \label{sec:SI}
Supplementary figures;
additional computational details and materials for the examples in the main text;
details of the model algorithms.
% \section*{Authors contributions}

\begin{acknowledgments}
This work received support from the National Natural Science Foundation of China (22220102001 and 22503081), the China Postdoctoral Science Foundation (2024M762886), and the CPSF Postdoctoral Fellowship Program (GZC20252381).
J.Z. and P.K thank Michele Parrinello, Enrico Trizio and Jaehoon Yang for discussions about this manuscript.
\end{acknowledgments}

%\clearpage
% Create the reference section using BibTeX:
\section*{Bibliography}
\bibliography{ref/main}
% \bibliographystyle{plain}

% begin supporting info
% reset counters for supporting info
\setcounter{section}{0}
\renewcommand{\thesection}{S\arabic{section}}
\setcounter{equation}{0}
\renewcommand{\theequation}{S\arabic{equation}}
\setcounter{figure}{0}
\renewcommand{\thefigure}{S\arabic{figure}}
\setcounter{table}{0}
\renewcommand{\thetable}{S\arabic{table}}

\clearpage
\onecolumngrid

{\Large\normalfont\sffamily\bfseries{{Supporting Information}}}
\pagenumbering{arabic}
\setcounter{page}{1}
\renewcommand{\thepage}{S\arabic{page}}

% start supporting info
\section{Computational details}

\subsection{Alanine dipeptide}

\subsubsection{Simulations setup}
We utilized the AMBER99-SB~\cite{ross2013} force field to describe the alanine dipeptide molecule in a vacuum.
The lengths of all chemical bonds involving hydrogen atoms were constrained at their equilibrium values using the RATTLE~\cite{andersen1983} algorithm.
The electrostatic and Lennard-Jones (LJ) interactions were described in a non-cutoff manner.
All MD propagations were carried out under the \textit{NVT} ensemble of 300 K using the \verb|OpenMM|~\cite{openmm} package with the Geodesic BAOAB Langevin integrator~\cite{leimkuhler2016} provided by \verb|OpenMMTools|.~\cite{openmmtools}
The integration time step and the friction coefficient were set to 2 \unit{\fs} and 1 \unit{\ps^{-1}}, correspondingly.
The number of geodesic drift steps was set to 4.

During the 20 \si{\ns} OPES simulations, we applied the committor $z$ variable as the CV.
For the biasing potential, we used a \texttt{BARRIER} parameter of 40 \unit{\kJ/\mole}, and the kernel functions were deployed every 500 steps.
During the 4 \si{\ns} TS-oriented Kolmogorov bias simulation, we selected a $\lambda$ parameter of about 1.2.

\subsubsection{Committor model training}

We regard the C7\textsubscript{eq} and C7\textsubscript{ax} states as the two boundary states.
Boundary conformations were obtained from a 10 \si{\ns} unbiased simulation of each state.
We adopted the dataset with 170,000 conformations reported in ref.~\citenum{kang2026committor} to train our committor model.
The dataset was randomly split into a training set (95\%) and a validation set (5\%).
The Pairformer architecture consisted of three layers, with 8 Gaussian basis functions, a pair representation dimensionality of 8, one attention head with a hidden dimensionality of 16, and a two-layer feedforward readout network.
For the optimization, we used the ADAM optimizer with an initial learning rate of 1E-4 modulated by an exponential decay with a multiplicative factor $\gamma$ of 0.99993. 
The $\alpha_1$ hyperparameter in the loss function was set to 10,000, and the $\alpha_2$ hyperparameter in the loss function was set to 1.
The training was performed for 1500 epochs with a batch size of 5,000.
The training loss curve is shown in Fig.~\ref{fig:s1}A.

\subsection{Chignolin mini-protein folding}

\subsubsection{Simulations setup}
We utilized the CHARMM22$^*$~\cite{piana2011robust} force field along with CHARMM TIP3P~\cite{mackerell1998tip} waters to study the folding and unfolding of chignolin (CLN025 peptide with a sequence of \verb|YYDPETGTWY|) in bulk solvent, sharing the same setup used for long unbiased simulations on this system~\cite{shaw2011fold} to have a direct comparison with those results.
The ASP and GLU residues, as well as the N- and C-terminus are simulated in their charged states.
The simulation box contains 1,907 water molecules, together with two sodium ions that neutralize the system.
All MD propagations were carried out under the \textit{NVT} ensemble of 340 K.
The electrostatic interactions were described using the Particle Mesh Ewald (PME) \cite{essmann1995} method with a real-space cutoff at 1.0 \unit{\nm}, and the Lennard-Jones (LJ) interactions were calculated with a distance cutoff at 1.0 \unit{\nm}.
The lengths of all chemical bonds involving hydrogen atoms were constrained at their equilibrium values.
The free energy calculations were carried out using the \verb|Gromacs| package~\cite{abraham2015gromacs} version 2024.3, with the SD Langevin integrator~\cite{kieninger2022sd} and the LINCS~\cite{hess1997lincs} constraint algorithm.
The integration time step and the friction coefficient were set to 2 \unit{\fs} and 1 \unit{\ps^{-1}}, correspondingly.
On the other hand, the shooting experiments were carried out with the \verb|OpenMM|~\cite{openmm} package with the Geodesic BAOAB Langevin integrator~\cite{leimkuhler2016} provided by \verb|OpenMMTools|.~\cite{openmmtools}
The integration time step and the friction coefficient were set to 2 \unit{\fs} and 10 \unit{\ps^{-1}}, correspondingly.
The number of geodesic drift steps was set to 2.

During the 2 \si{\us} OPES simulations, we applied the LASSO approximation to the committor $z$ variable as the CV (see the below sections for detail).
For the biasing potential, we used a \texttt{BARRIER} parameter of 30 \unit{\kJ/\mole}, and the kernel functions were deployed every 500 steps.
During the 20 \si{\ns} TS-oriented Kolmogorov bias simulations, we selected a $\lambda$ parameter of about 1.2.
To focus on sampling the transition conformations ($q(\symbf{x}) \approx 0.5$), we used a truncated Kolmogorov bias expression instead of the original equation:
\begin{equation}
V^T_{\mathcal{K}}(\symbf{x})=-\frac{\lambda \symsf{cos}^2\left(\pi(q(\symbf{x})-0.5)\right)}{\beta}\symsf{log}\left|\nabla_{\symbf{u}}q(\symbf{x})\right|^2.
\label{equ:k_bias}
\end{equation}

\subsubsection{Committor model training}

\begin {table}[th!]
\caption {Sensitivity analysis results of chignolin} \label{tab:s1}
\begin{center}
\begin{tabular}{ |c|c| } 
\hline
Atom pair & Normalized sensitivity \\ 
\hline
TYR1N-TYR10O  & 1.00 \\
ASP3O-GLY7O   & 0.82 \\
TYR1O-TYR10O  & 0.74 \\
TYR1N-TYR10C  & 0.73 \\
TYR1N-GLU5O   & 0.63 \\
TYR1O-TRP9O   & 0.63 \\
TYR1Ca-TYR10O & 0.62 \\
TYR1N-TYR10Ca & 0.61 \\
TYR2O-TYR10O  & 0.61 \\
TYR2N-TYR10O  & 0.59 \\
TYR1N-TYR10N  & 0.59 \\
TYR1N-TRP9O   & 0.56 \\
TYR1N-THR6O   & 0.52 \\
GLU5O-TYR10O  & 0.51 \\
ASP3O-THR8N   & 0.51 \\
TYR2N-TYR10C  & 0.51 \\
TYR2N-THR6O   & 0.50 \\
GLU5N-TYR10O  & 0.49 \\
TYR1N-THR8O   & 0.49 \\
TYR2O-THR6O   & 0.48 \\
GLU5N-TRP9O   & 0.48 \\
TYR2O-TRP9O   & 0.47 \\
\hline
\end{tabular}
\end{center}
\end {table}

We regard the folded and completely unfolded states as the two boundary states.
Boundary conformations were obtained from a 10 \si{\ns} unbiased simulation of each state.
Specifically, for the simulation of the unfolded states, we restrained the end-to-end distance of the peptide to be longer than 1.5 \si{\nm}, the alpha carbon RMSD to be larger than 0.5 \si{\nm} and the distance between ASP3N and THR8O to be longer than 1.2 \si{\nm}.
We then adopted the dataset with about 370,000 conformations reported in ref.~\citenum{trizio2025committor} to train our committor model.
The dataset was randomly split into a training set (95\%) and a validation set (5\%).
The Pairformer architecture consisted of two layers, with 16 Gaussian basis functions, a pair representation dimensionality of 8, one attention head with a hidden dimensionality of 16, and a two-layer feedforward readout network.
For the optimization, we used the ADAM optimizer with an initial learning rate of 5E-5 modulated by an exponential decay with a multiplicative factor $\gamma$ of 0.99997. 
The $\alpha_1$ hyperparameter in the loss function was set to 10,000, and the $\alpha_2$ hyperparameter in the loss function was set to 1.
The training was performed for 2000 epochs with a batch size of 3,000.
The training loss curve is shown in Fig.~\ref{fig:s1}B.

\begin{figure}[th!]
\centering
\includegraphics[width=\linewidth]{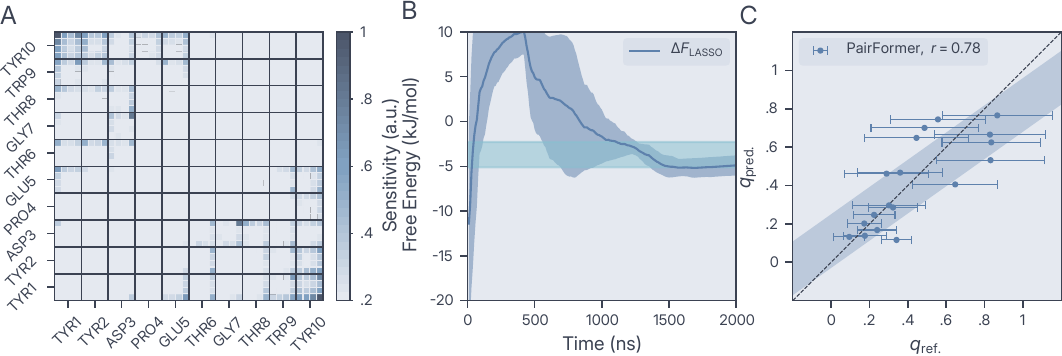}
\caption{Additional results of the chignolin folding process. (A) Sensitivity score of each  backbone atom pair. As illustrated, interatomic distances between residues ASP3, THR6, GLY7, and THR8 contribute significantly to the committor. (B) Free energy difference between the folded and unfolded state during the OPES simulations. The average value estimated from three independent simulations are shown as the blue solid line, whereas the uncertainty, computed as the standard deviation over the replicas, is depicted as the shaded blue region. The light blue region represents the $\pm 0.5 k_BT$ range of the reference free energy, which is obtained from ref.~\citenum{shaw2011fold}. (C) Performance of the committor model tested on an independent set of conformations. Reference values and their associated uncertainties were calculated from three groups of 20 independent shots for each conformation. The shaded blue region represents the linear fit between the reference and predicted values, with the boundaries denoting $\pm$ the standard error of the fitting.}
\label{fig:s2}
\end{figure}

After fitting the committor, we carried out shooting experiments on 18 untrained random conformations to examine the accuracy of the model (Fig.~\ref{fig:s2}C).
For each conformation, three groups of 20 independent shots were performed under a large friction coefficient of 10 \unit{\ps^{-1}}.
We then used bootstrapping with MC sampling to calculate the error of the reference committor values.

\subsubsection{LASSO analysis}
Table~\ref{tab:s1} lists the important interatomic distances found by the sensitivity analysis.
Using such distances, the LASSO analysis found the following approximation to the committor $z$:
\begin{equation}
\begin{split}
z(\symbf{x})\ &\propto 0.53*d_{TYR1O-TRP9O}+0.40*d_{ASP3O-THR8N}+0.21*d_{TYR2N-TYR10C} \\
& +0.21*d_{TYR1N-THR8O}+0.12*d_{TYR1O-TYR10O}-0.11*d_{GLU5O-TYR10O} \\
& +0.02*d_{ASP3O-GLY7O}.
\label{equ:chignolin_lasso}
\end{split}
\end{equation}

\subsection{Calixarene host-guest systems}

\subsubsection{Simulations setup}
We utilized the Amber General Force Field (GAFF)~\cite{wang2004development} with the RESP~\cite{bayly1993well} charges and the TIP3P~\cite{jorgensen1983tip3p} waters to describe the calixarene system in bulk solvent.
For \ce{G2}, \ce{G5} and \ce{G6} ligands, a pair of calixarene host molecule and ligand molecule was solvated at a density of about 1 \unit{g/cm^3} with about 2110 water molecules in a cubic box of 4.03 $\times$ 4.03 $\times$ 4.03 \unit{\nm^3}.
Extra sodium ions are included to counterbalance excess charges. 
All MD propagations were carried out under the \textit{NVT} ensemble of 300 K using the SD Langevin integrator~\cite{kieninger2022sd}.
The electrostatic interactions were described using the Particle Mesh Ewald (PME) \cite{essmann1995} method with a real-space cutoff at 1.0 \unit{\nm}, and the Lennard-Jones (LJ) interactions were calculated with a distance cutoff at 1.0 \unit{\nm}.
The lengths of all chemical bonds were constrained at their equilibrium values using the LINCS~\cite{hess1997lincs} algorithm.
The free energy calculations were carried out using the \verb|Gromacs| package~\cite{abraham2015gromacs} version 2024.3.
The integration time step and the friction coefficient were set to 2 \unit{\fs} and 1 \unit{\ps^{-1}}, correspondingly.
The shooting calculations were carried out using the \verb|Gromacs| package~\cite{abraham2015gromacs} version 2025.4.
The integration time step and the friction coefficient were set to 2 \unit{\fs} and 10 \unit{\ps^{-1}}, correspondingly.

In our simulations, we used a funnel restraint~\cite{limongelli2013funnel} equivalent to the one previously employed by ref.\citenum{rizzi2021water} on the same system.
Here, we summarize the details of such a restraint, while more details can be found in the original works.
The funnel limits the space available to the ligand in the unbound state U by confining it to a cylindrical volume above the binding site. 
As the ligand approaches the binding site, the funnel restraint becomes wider so that its presence does not affect the binding process itself. 
Having aligned the system with \verb|plumed|~\cite{plumed} to a reference configuration where the binding axis is found along the vertical axis, we defined $h$ as the projection on the binding axis of the center of the carbon atoms of each ligand and $r$ as its radial component.
When $h>1$ \si{\nm}, the funnel surface is a cylinder with a radius $R_{\symsf{cyl}} = 0.2$ \si{\nm} with its axis along the vertical direction. 
When $h<1$ \si{\nm}, the funnel opens into an umbrella-like shape with a 45-degree angle whose surface is defined by $r=1.2-h$.
The force that, for a displacement $x$, pushes the ligand away from the funnel’s surface is harmonic $-k_Fx$ with $k_F=2000$ \si{\kJ/\mole/\nm^2}. 
A further harmonic restraint is applied on $h$ to prevent the ligand from getting too far from the host, reaching the upper boundary of the simulation box. 
The corresponding force is $-k_U(h-1.8)$ for $h > 1.8$ \si{\nm} and $k_U=4000$ \si{\kJ/\mole/\nm^2}.
During training, we applied a static ``cap'' potential on $h$ to further enhance the sampling of the transition region.
The corresponding force is $-k_C$ for $h > 1$ \si{\nm} and $k_C=25$ \si{\kJ/\mole/\nm}.

Because of the funnel presence, the free energy difference between the bound and the true unbound state that we extract from enhanced sampling simulations needs a correction that can be calculated as:
\begin{equation}
\Delta G = -\frac{1}{\beta} \symsf{log} \left( C_0 \pi R_{\symsf{cyl}}^2 \int_B dh \symsf{exp}\left[-\beta \left(W(h) - W_U\right)\right] \right)
\end{equation}
where $\beta = 1/(k_BT)$, $C_0$ = 1/1.66 $\si{\nm}^{-3}$ is the standard concentration, $h$ is the coordinate along the funnel’s axis, $W(h)$ is the free energy along the funnel axis and $W_U$ its reference value in state U. 
More precisely, we defined $W_U$ as the average free energy value in the interval 1.6 $\si{\nm}$ < $h$ < 1.8  $\si{\nm}$.
The integral is computed over the state B region that we define as 0.3 $\si{\nm}$ < $h$ < 0.8  $\si{\nm}$.

During the 200 \si{\ns} final round OPES simulations, we applied the committor $z$ variable and the $h$ variable as the CVs.
For the biasing potential, we used a \texttt{BARRIER} parameter of 45 \unit{\kJ/\mole}, and the kernel functions were deployed every 500 steps.
During the 10 \si{\ns} final round TS-oriented Kolmogorov bias simulation, we selected a $\lambda$ parameter of about 1.2.

\subsubsection{Committor model training}
\begin{figure}[th!]
\centering
\includegraphics[width=\linewidth]{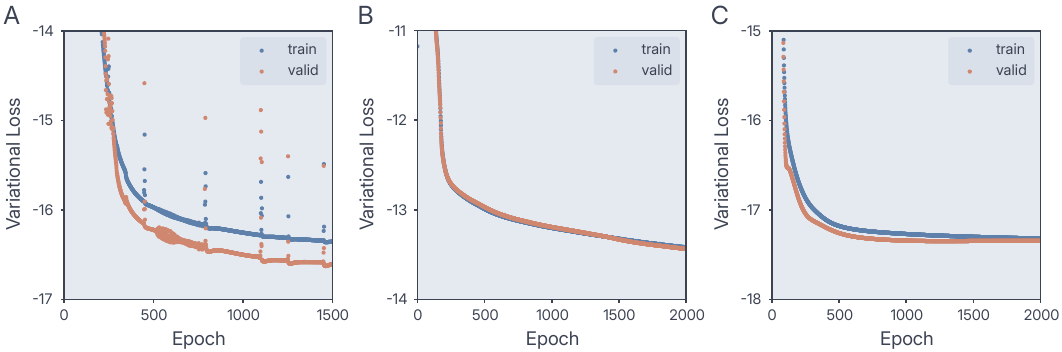}
\caption{Training behavior of each system in the final training iteration. The blue curve shows the variational loss evaluated on the training set, while the variational loss evaluated on the training set is shown by the orange curve. (A) Alanine dipeptide. (B) Chignolin. (C) Calixarene host-guest systems.}
\label{fig:s1}
\end{figure}

In contrast to the above two examples, we performed the self-consistent iterations to build the committor model for the calixarene host-guest systems.
We regard the dry bound state and wet unbound states as the two boundary states.
Boundary conformations were obtained from a 5 \si{\ns} unbiased simulation for each state of each system.
Specifically, for the simulation of the unfolded states, we restrained the ligand-pocket distance $h$ to be longer than 1.6 \si{\nm} but shorter than 1.8 \si{\nm}.
The Pairformer architecture consisted of two layers, with 16 Gaussian basis functions, a pair representation dimensionality of 8, one attention head with a hidden dimensionality of 16, and a two-layer feedforward readout network.
Besides, geometric centers of the three atom groups H\textsubscript{B},  H\textsubscript{M} and  H\textsubscript{T} were selected as centers of the solvent embedding, which uses the above parameters: $NN = 2$, $MM = 6$, $r_0 = 0.25$, $d_0 = 1.0$.
Water oxygen atoms within 1 \si{\nm} of the centers were considered when calculating the coordination numbers.
For the optimization, we used the ADAM optimizer with an initial learning rate of 5E-5 modulated by an exponential decay with a multiplicative factor $\gamma$ of 0.99997. 
The $\alpha_1$ hyperparameter in the loss function was set to 10,000, and the $\alpha_2$ hyperparameter in the loss function was set to 1.
The training was performed for 2000 epochs with a batch size of 3,000.
The training loss curve is shown in Fig.~\ref{fig:s1}C.

As proposed in ref.~\cite{kang2026committor}, we did not start the iterations from committor trained solely using the boundary term.
Instead, we performed 200 \si{\ns} OPES simulations for each system using suboptimal CVs, namely the $h$ variable and the cosine of the angle between the ligands and the binding axis.~\cite{rizzi2021water}
For these simulations, we used a \texttt{BARRIER} parameter of 50 \unit{\kJ/\mole}, and the kernel functions were deployed every 500 steps.
Based on the boundary conformations and the primary OPES data, we carried out two fitting iterations.
In Table~\ref{tab:s2} we listed in each iteration the corresponding dataset size, the applied biasing potential parameters, the simulation time $t_s$ and the lowest value obtained for the functional $\mathcal{K}_m$, which provides a quality and convergence measure.
\begin {table}[h!]
\caption {Summary of the iterative procedure for the calixarene host-guest systems.} \label{tab:s2}
\begin{center}
\begin{tabular}{ |c|c|c|c|c|c| } 
\hline
Iteration & Dataset size & $\symsf{log}{\mathcal{K}}_m$ (a.u.) & OPES \texttt{BARRIER} (\si{\kJ/\mole}) & $\lambda$ & $t_s$ \\ 
\hline
0 & 140000 & -15.97  & 50 &   -  & 2 * 3 * 5 \si{\ns} + 3 * 200 \si{\ns} \\
1 & 380000 & -17.30  & 40 & 0.6  & 3 * 200 \si{\ns}  \\
\hline
\end{tabular}
\end{center}
\end {table}

\begin{figure}[th!]
\centering
\includegraphics[width=\linewidth]{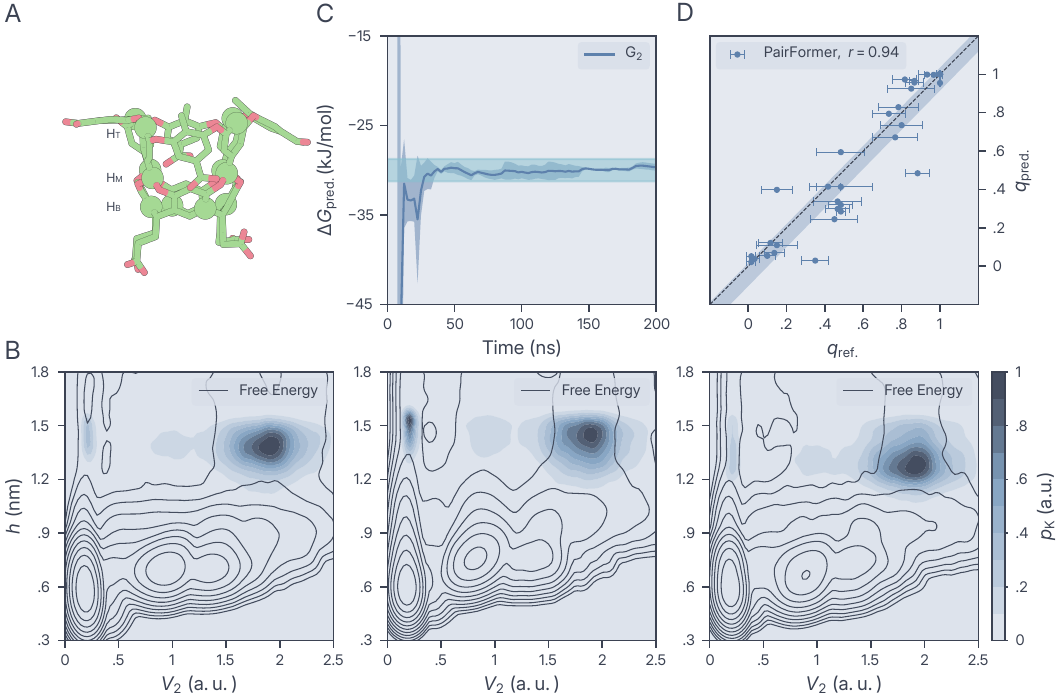}
\caption{Calixarene host-guest systems. (A) Positions of the three atom groups used to build inputs of the model. The 12 selected atoms are shown in spheres. (B) Contour plots of the unnormalized Kolmogorov distribution $e^{-\beta U_{\mathcal{K}}(\symbf{x})}$ of each system projected in subspace spanned by the water coordination number of a virtual point inside the binding cavity (V\textsubscript{2}) and the projection of the ligand molecule on the binding axis ($h$). Grey isolines in the plots indicate the free energy. From left to right, the three panels show the plot of the G\textsubscript{2}, G\textsubscript{5} and G\textsubscript{6} ligands, respectively. (C) Free energy difference between the bound and unbound state after the funnel correction of the G\textsubscript{2} ligand during the OPES simulations. The average value estimated from three independent simulations are shown as the blue solid line, whereas the uncertainty, computed as the standard deviation over the replicas, is depicted as the shaded blue region. The light blue region represents the $\pm 0.5 k_BT$ range of the reference free energy, which is obtained from ref.~\citenum{trizio2025committor}. (D) Performance of the committor model tested on independent sets of conformations of each complex. Reference values and their associated uncertainties were calculated from three groups of 20 independent shots for each conformation. The shaded blue region represents the linear fit between the reference and predicted values, with the boundaries denoting $\pm$ the standard error of the fitting.}
\label{fig:s3}
\end{figure}

After fitting the committor, we carried out shooting experiments on 12 untrained random conformations for each system to examine the accuracy of the model (Fig.~\ref{fig:s3}D).
For each conformation, three groups of 20 independent shots were performed under a large friction coefficient of 10 \unit{\ps^{-1}}.

\clearpage

\section{Details about the model architecture}

\begin{algorithm}
\caption{Triangular gated self-attention around starting node}
\begin{algorithmic}[1]
\Statex \textbf{def} $\mathsf{TriangleAttentionStartingNode} \left(\left\{\symbf{p}_{ij}\right\},c,N_{head}\right)$\hfill$\symbf{p}_{ij}\in\mathbb{R}^c$:
%\Statex \Comment{\# Input projections}
\State $\symbf{p}_{ij} \leftarrow \symsf{LayerNorm}\left(\symbf{p}_{ij}\right)$
\State $\symbf{q}_{ij}^h, \symbf{k}_{ij}^h, \symbf{v}_{ij}^h = \symsf{LinearNoBias}\left(\symbf{p}_{ij}\right)$ \hfill $\symbf{q}_{ij}^h, \symbf{k}_{ij}^h, \symbf{v}_{ij}^h \in \mathbb{R}^c,\ h \in \left\{1, \dots, N_{head}\right\}$
\State $\symbf{b}_{ij}^h = \symsf{LinearNoBias}\left(\symbf{p}_{ij}\right)$
\State $\symbf{g}_{ij}^h = \symsf{sigmoid}\left(\symsf{Linear}\left(\symbf{p}_{ij}\right)\right)$ \hfill $\symbf{g}_{ij}^h \in \mathbb{R}^c$
%\Statex \Comment{\# Attention}
\State $\symbf{a}_{ijk}^h = \symsf{softmax}_k \left( \frac{1}{\sqrt{c}} {\symbf{q}_{ij}^{h}}^{\top} \symbf{k}_{ik}^h + \symbf{b}_{jk}^h \right)$
\State $\symbf{o}_{ij}^h = \symbf{g}_{ij}^h \odot \sum_k \symbf{a}_{ijk}^h \symbf{v}_{ik}^h$
%\Statex \Comment{\# Output projection}
\State $\tilde{\symbf{p}}_{ij} = \symsf{Linear}\left(\symsf{concat}_h\left(\symbf{o}_{ij}^h\right)\right)$ \hfill $\tilde{\symbf{p}}_{ij} \in \mathbb{R}^{c}$
\State \textbf{return} $\left\{\tilde{\symbf{p}}_{ij}\right\}$
\end{algorithmic}
\end{algorithm}

\begin{algorithm}
\caption{Triangular gated self-attention around ending node}
\begin{algorithmic}[1]
\Statex \textbf{def} $\mathsf{TriangleAttentionEndingNode}\left(\left\{\symbf{p}_{ij}\right\},c,N_{head}\right)$\hfill$\symbf{p}_{ij}\in\mathbb{R}^c$:
% \Statex \Comment{\# Input projections}
\State $\symbf{p}_{ij} \leftarrow \symsf{LayerNorm}\left(\symbf{p}_{ij}\right)$
\State $\symbf{q}_{ij}^h, \symbf{k}_{ij}^h, \symbf{v}_{ij}^h = \symsf{LinearNoBias}\left(\symbf{p}_{ij}\right)$ \hfill $\symbf{q}_{ij}^h, \symbf{k}_{ij}^h, \symbf{v}_{ij}^h \in \mathbb{R}^c,\ h \in \left\{1, \dots, N_{head}\right\}$
\State $\symbf{b}_{ij}^h = \symsf{LinearNoBias}\left(\symbf{p}_{ij}\right)$
\State $\symbf{g}_{ij}^h = \symsf{sigmoid}\left(\symsf{Linear}\left(\symbf{p}_{ij}\right)\right)$ \hfill $\symbf{g}_{ij}^h \in \mathbb{R}^c$
% \Statex \Comment{\# Attention}
\State $\symbf{a}_{ijk}^h = \symsf{softmax}_k \left( \frac{1}{\sqrt{c}} {\symbf{q}_{ij}^{h}}^{\top} \symbf{k}_{kj}^h + \symbf{b}_{ki}^h \right)$
\State $\symbf{o}_{ij}^h = \symbf{g}_{ij}^h \odot \sum_k \symbf{a}_{ijk}^h \symbf{v}_{kj}^h$
% \Statex \Comment{\# Output projection}
\State $\tilde{\symbf{p}}_{ij} = \symsf{Linear}\left(\symsf{concat}_h\left(\symbf{o}_{ij}^h\right)\right)$ \hfill $\tilde{\symbf{p}}_{ij} \in \mathbb{R}^{c}$
\State \textbf{return} $\left\{\tilde{\symbf{p}}_{ij}\right\}$
\end{algorithmic}
\end{algorithm}

\begin{algorithm}
\caption{Transition layer}
\begin{algorithmic}[1]
\Statex \textbf{def} $\mathsf{Transition}\left(\left\{\symbf{p}_{ij}\right\},t=4\right)$\hfill$\symbf{p}_{ij}\in\mathbb{R}^c$:
\State $\symbf{p}_{ij}\leftarrow\symsf{LayerNorm}\left(\symbf{p}_{ij}\right)$
\State $\symbf{a}_{ij} = \symsf{LinearNoBias}\left(\symbf{p}_{ij}\right)$ \hfill $\symbf{a}_{ij}\in\mathbb{R}^{t\cdot c}$
\State $\symbf{b}_{ij} = \symsf{LinearNoBias}\left(\symbf{p}_{ij}\right)$ \hfill $\symbf{b}_{ij}\in\mathbb{R}^{t\cdot c}$
\State $\symbf{p}_{ij}\leftarrow\symsf{LinearNoBias}\left(\symsf{swish}\left(\symbf{a}_{ij}\right)\odot\symbf{b}_{ij}\right)$ \hfill $\symbf{p}_{ij}\in\mathbb{R}^c$
\State \textbf{return} $
\left\{\symbf{p}_{ij}\right\}$
\end{algorithmic}
\end{algorithm}

\begin{algorithm}
\caption{Pairformer block with pair channels only}
\begin{algorithmic}[1]
\Statex \textbf{def} $\mathsf{PairformerBlock}\left(\left\{\symbf{p}_{ij}\right\}\right)$\hfill$\symbf{p}_{ij}\in\mathbb{R}^c$:
\State $\left\{\symbf{p}_{ij}\right\} \leftarrow \left\{\symbf{p}_{ij}\right\} + \symsf{TriangleAttentionStartingNode}\left(\left\{\symbf{p}_{ij}\right\}\right)$
\State $\left\{\symbf{p}_{ij}\right\} \leftarrow \left\{\symbf{p}_{ij}\right\} + \symsf{TriangleAttentionEndingNode}\left(\left\{\symbf{p}_{ij}\right\}\right)$
\State $\left\{\symbf{p}_{ij}\right\} \leftarrow \left\{\symbf{p}_{ij}\right\} + \symsf{Transition}\left(\left\{\symbf{p}_{ij}\right\}\right)$
\State \textbf{return} $\left\{\symbf{p}_{ij}\right\}$
\end{algorithmic}
\end{algorithm}

\begin{algorithm}
\begin{algorithmic}[1]
\caption{Coordination number model}
\Statex \textbf{def} $\mathsf{CNModel}\left(\left\{\symbf{x}^E_i\right\},\left\{\symbf{x}^C_i\right\},NN,MM,r_0,d_0,d_{max}\right)$\hfill$\symbf{x}^E_i,\symbf{x}^C_i\in\mathbb{R}^3$:
\State $\symbf{d}_{ij} = \left(\left\Vert\symbf{x}^C_i-\symbf{x}^E_j\right\Vert-d_0\right)/r_0$
\State $c_i = \sum_j\left(1-\left(\symbf{d}_{ij}\right)^{NN}\right)/\left(1-\left(\symbf{d}_{ij}\right)^{MM}\right)$\hfill$c_i\in\mathbb{R}$
\State $d_{max} \leftarrow \left(d_{max}-d_0\right)/r_0$
\State $c_{max} = \left(1-\left(d_{max}\right)^{NN}\right)/\left(1-\left(d_{max}\right)^{MM}\right)$
\State $c_i \leftarrow \left(c_i-c_{max}\right)/\left(1-c_{max}\right)$
\State $\symbf{c} = \symsf{Linear}\left(\left\{c_i\right\}\right)$\hfill$\symbf{c}\in\mathbb{R}^{2\cdot\left|\left\{\symbf{x}^C_i\right\}\right|}$
\State \textbf{return} $\symbf{c}$
\end{algorithmic}
\end{algorithm}

\begin{algorithm}
\caption{Pairformer-based committor model}
\begin{algorithmic}[1]
\Statex \textbf{def} $\mathsf{PairformerCommittor}\left(\left\{\symbf{x}^S_i\right\},\left\{\symbf{x}^E_i\right\},\left\{\symbf{x}^C_i\right\},\left\{\symbf{t}^A_i\right\},\left\{\symbf{t}^R_i\right\},N_{block}\right)$\hfill$\symbf{x}^S_i,\symbf{x}^E_i,\symbf{x}^C_i\in\mathbb{R}^3,\ \symbf{t}^A_i,\symbf{t}^R_i\in\mathbb{R}$:
\State $\symbf{d}_{ij} = \left\Vert\symbf{x}^S_i-\symbf{x}^S_j\right\Vert $
\State $\symbf{b}_{ij} = \symsf{LinearNoBias}\left(\symsf{RBF}\left(\symbf{d}_{ij}\right) \odot \symsf{Decay}\left(\symbf{d}_{ij}\right)\right)$\hfill$\symbf{b}_{ij}\in\mathbb{R}^c$
\State $\left\{\symbf{p}_{ij}\right\} = \mathsf{PairformerBlock}\left(\left\{\symbf{b}_{ij}\right\}\right)$\hfill$\symbf{p}_{ij}\in\mathbb{R}^c$
\State $\symbf{t}_{i} = \symsf{concat}\left(\symsf{OneHot}\left(\symbf{t}^A_i\right),\symsf{OneHot}\left(\symbf{t}^R_i\right)\right)$\hfill$\symbf{t}_{i}\in\mathbb{R}^c$
\State $\symbf{p}_{ij} \leftarrow \symsf{Linear}\left(\symsf{concat}\left(\symbf{t}_{i},\symbf{p}_{ij},\symbf{t}_{j}\right)\right)$\hfill$\symbf{p}_{ij}\in\mathbb{R}^c$
\For{$l \in \left[2, \dots, N_{block}\right]$}
\State $\left\{\symbf{p}_{ij}\right\} \leftarrow \symsf{PairformerBlock}\left(\left\{\symbf{p}_{ij}\right\}\right)$
\EndFor
\State $\symbf{z} = \symsf{concat}\left(\symsf{AvgPool2D}_{ij}\left(\left\{\symbf{p}_{ij}\right\}\right),\symsf{CNModel}\left(\left\{\symbf{x}^E_i\right\},\left\{\symbf{x}^C_i\right\}\right)\right)$\hfill$\symbf{z}\in\mathbb{R}^c$
\State $z = \symsf{Linear}\left(\symsf{ShiftedSoftplus}\left(\symsf{Linear}\left(\symbf{z}\right)\right)\right)$\hfill$z\in\mathbb{R}$
\State $q = \symsf{sigmoid}\left(z\right)$\hfill$q\in\mathbb{R}$
\State \textbf{return} $\left\{z,q\right\}$
\end{algorithmic}
\end{algorithm}

\clearpage

\section{Supplementary figures}

\clearpage
% end supporting info

% \clearpage
% \bibliography{ref/main}
\end{document}